\documentclass[final,3p,times,authoryear]{elsarticle}
\usepackage{amssymb}
\usepackage{amsmath}
\usepackage{lineno}
\usepackage{graphicx}
\usepackage{caption}
\usepackage{makecell}
\usepackage{hyperref}
\hypersetup{colorlinks=true,urlcolor=blue}

\journal{Communications in Nonlinear Science and Numerical Simulation}

\begin{document}

\begin{frontmatter}

\title{A Lie-Group Approach to Alfv\'enic Turbulence: The SO(3) Rotational p-model}

\author[aff1]{Daniele Telloni\corref{cor1}}
\author[aff2,aff1]{Sabrina Guastavino}
\author[aff3,aff4]{Gary P. Zank}
\author[aff5,aff6]{Luca Sorriso-Valvo}
\author[aff2,aff1]{Michele Piana}

\affiliation[aff1]{organization={National Institute for Astrophysics, Astrophysical Observatory of Torino},
                        addressline={Via Osservatorio 20},
                        city={Pino Torinese},
                        postcode={10025},
                        country={Italy}}
\affiliation[aff2]{organization={Department of Mathematics, University of Genoa},
                        addressline={Via Dodecaneso 35},
                        city={Genoa},
                        postcode={16146},
                        country={Italy}}
\affiliation[aff3]{organization={Center for Space Plasma and Aeronomic Research, University of Alabama in Huntsville},
                        city={Huntsville},
                        postcode={35805},
                        state={AL},
                        country={USA}}
\affiliation[aff4]{organization={Department of Space Science, University of Alabama in Huntsville},
                        city={Huntsville},
                        postcode={35805},
                        state={AL},
                        country={USA}}
\affiliation[aff5]{organization={National Research Council, Institute for Plasma Science and Technology},
                        addressline={Via Amendola 122/D},
                        city={Bari},
                        postcode={70126},
                        country={Italy}}
\affiliation[aff6]{organization={Department of Electromagnetics and Plasma Physics, School of Electrical Engineering and Computer Science, KTH Royal Institute of Technology},
                        city={Stockholm},
                        postcode={11428},
                        country={Sweden}}

\cortext[cor1]{Corresponding author: Daniele Telloni (\href{daniele.telloni@inaf.it}{daniele.telloni@inaf.it})}

\begin{abstract}
	The inner heliosphere is permeated by fully developed, highly intermittent nonlinear turbulence. Conventional stochastic models synthesize these multiscale fluctuations by injecting additive noise within a flat Euclidean space. However, such approaches fundamentally violate the topological constraints of ideal magnetohydrodynamics, inadvertently breaking the kinetic incompressibility of the plasma and introducing spurious magnetic compressibility. To preserve the exact geometrical invariants of Alfv\'enic fluctuations, a novel mathematical framework is presented: the rotational p-model. By mapping a multifractal singular measure directly into the Lie algebra $\mathfrak{so}(3)$ and integrating the associated stochastic differential equations in the Stratonovich sense on the spherical manifold $S^{2}$, the model generates highly intermittent, scale-dependent rotational dynamics. This Lie-group formulation conserves the magnetic field modulus while perfectly reproducing the fractional spectral envelopes and the non-Gaussian directional intermittency observed in space plasmas. Furthermore, by isolating purely directional reorientations, the framework provides exact analytical control over the waiting time statistics of magnetic intermittent bursts and the multifractal scaling exponents of angular deflections, offering a powerful new paradigm for the study of nonlinear turbulent phenomena.
\end{abstract}

\begin{highlights}
	\item Multifractal cascades mapped into $\mathfrak{so}(3)$ generate non-Gaussian angular dynamics.
	\item Stratonovich SDE integration on $S^{2}$ ensures strict magnetic incompressibility.
	\item The framework exactly isolates purely directional intermittency and waiting times.
\end{highlights}

\begin{keyword}
	Alfv\'enic turbulence \sep Lie groups \sep Multifractal cascades \sep Stochastic differential equations \sep Interplanetary magnetic field
	\PACS 02.20.Sv \sep 02.20.Tw \sep 02.60.-x \sep 05.10.-a \sep 52.30.Cv \sep 52.35.Ra \sep 94.05.Lk \sep 95.75.Pq \sep 96.50.Ci
	\MSC[2020] 22E70 \sep 65C30 \sep 65T50 \sep 76F55 \sep 76M35 \sep 76M60 \sep 76W05
\end{keyword}

\end{frontmatter}


\section{Introduction}
\label{sec:introduction}
The inner heliosphere behaves as a highly complex, collisionless, and magnetized fluid exhibiting fully developed nonlinear turbulence. From a macroscopic perspective, the solar wind expanding through the interplanetary medium possesses a distinct dual nature: a slow, dense, and highly compressible wind typically confined to the equatorial streamer belt, and a fast, tenuous wind originating from polar coronal holes. The fast solar wind is characteristically dominated by high Alfv\'enicity, meaning that it is permeated by a continuum of outward-propagating Alfv\'en wave packets \citep{2013LRSP...10....2B}. In this highly Alfv\'enic and quite idealized regime, the plasma obeys a strict restriction of kinematic incompressibility, mathematically articulated as $\nabla\cdot\delta\mathbf{V}=0$, where $\delta\mathbf{V}$ denote plasma velocity field fluctuations. Alfv\'en waves are purely transverse perturbations of the magnetic field lines; the plasma oscillates in a direction strictly perpendicular to both the wave propagation vector $\mathbf{k}$ and the background magnetic field $\mathbf{B}_{0}$. In the Fourier domain, the spatial divergence operator $(\nabla\cdot)$ translates into the scalar product with the wave vector $(i\mathbf{k})$. Since the velocity perturbation $\delta\mathbf{V}$ is geometrically orthogonal to the propagation direction $\mathbf{k}$, their scalar product is identically zero ($\mathbf{k}\cdot\delta\mathbf{V}=0$). Applying the inverse transformation to the physical space rigorously yields $\nabla\cdot\delta\mathbf{V}=0$.

Inserting this zero-divergence condition into the mass continuity equation ($\partial\rho/\partial t+\nabla\cdot(\rho\mathbf{V})=0$) demonstrates that the local plasma density remains constant ($\delta\rho=0$). Unlike magnetoacoustic waves, the fluid does not undergo longitudinal compressions or rarefactions, experiencing exclusively lateral shear stresses.

This kinematic behavior profoundly dictates the magnetic topology. Alfv\'en's frozen-in flux theorem \citep{1942Natur.150..405A} states that, in an ideal plasma characterized by infinite electrical conductivity ($\sigma\rightarrow\infty$), magnetic field lines are topologically constrained to move coherently with the fluid elements. Consequently, the magnetic flux $\Phi_{B}$ passing through any open surface $S(t)$ bounded by a closed contour $C(t)$ moving covariantly with the plasma is a temporal invariant. The integral formulation of the theorem asserts that the total derivative of the magnetic flux is strictly zero:

\begin{equation}
	\frac{d\Phi_{B}}{dt}=\frac{d}{dt}\int_{S(t)}\mathbf{B}\cdot d\mathbf{S}=0.
\end{equation}

By exploiting the Reynolds transport theorem, the Maxwell-Faraday equation ($\nabla\times\mathbf{E}=-\partial\mathbf{B}/\partial t$) and Ohm's law for ideal magnetohydrodynamics ($\mathbf{E}+\mathbf{V}\times\mathbf{B}=0$), the frozen-in flux condition is expressed in the differential form of the induction equation:

\begin{equation}
	\frac{\partial\mathbf{B}}{\partial t}=\nabla\times(\mathbf{V}\times\mathbf{B}).
\end{equation}

In this formulation, the magnetic diffusion term ($\eta\nabla^{2}\mathbf{B}$) is absent. The evolution of the $\mathbf{B}$ field is dictated exclusively by the convective term: the plasma velocity field $\mathbf{V}$ transports, stretches, and deflects the magnetic field lines without breaking or reconnecting them, thereby preserving the initial magnetic topology at any given instant $t$. Since the Alfv\'enic plasma velocity field is devoid of compressive gradients ($\nabla\cdot\delta\mathbf{V}=0$), the frozen-in magnetic field lines are uniquely deflected spatially and never locally squeezed or rarefied. This immunity to geometric compression mathematically guarantees the absence of perturbations in the local magnetic field modulus ($\delta|\mathbf{B}|=0$).

Despite these stringent topological constraints, conventional synthetic modeling of the interplanetary magnetic field $\mathbf{B}(t)$ relies on the injection of spatially homogeneous stochastic perturbations within the Euclidean space $\mathbb{R}^{3}$. In the linear paradigm, the total field is expressed as the superposition of a background field $\mathbf{B}_{0}(t)$ and a fractal fluctuation $\delta\mathbf{B}(t)$:

\begin{equation}
	\mathbf{B}(t)=\mathbf{B}_{0}(t)+\delta\mathbf{B}(t).
\end{equation}

Vector addition within a flat Euclidean space induces a spontaneous symmetry breaking of this incompressibility:

\begin{equation}
	|\mathbf{B}(t)|^{2}=|\mathbf{B}_{0}|^{2}+|\delta\mathbf{B}|^{2}+2(\mathbf{B}_{0}\cdot\delta\mathbf{B}).
\end{equation}

Even when rigorous orthogonality is imposed (i.e., $\mathbf{B}_{0}\cdot\delta\mathbf{B}=0$), the vector norm undergoes a quadratic deviation $\Delta|\mathbf{B}|^{2}\sim\mathcal{O}(|\delta\mathbf{B}|^{2})$, introducing spurious magnetic compressibility.

Early attempts to build solenoidal and incompressive magnetic fluctuations were able to reduce the fluctuations' amplitude to a few percent \citep{2012PhRvL.109w1102R}, which was more recently further reduced to less than $1\%$ \citep{2019ApJ...881L...5V}. Those analytical models generated strong local discontinuities, which locally break the requirement of constant field magnitude and undermine the exploitability of the fields. In another attempt, \citet{2022JPlPh..88e1703S} built a method to amplify initially low-amplitude Alfv\'enic fluctuations into large-amplitude ones, preserving continuity yet developing tendency to generate discontinuities. These fluctuations, however, were not shown to reproduce solar wind characteristics such as spectral decay and intermittency. More recently, \citet{2026ApJ..1007L..18T} built a model able to reproduce several solar wind properties, which again unavoidably generated localized discontinuities, yet providing subsets of incompressible fluctuations. In that case, discontinuities were compared to solar wind magnetic switchbacks, namely large-amplitude Alfv\'enic rotations, often observed with a full reversal of the radial component \citep{2026SSRv..222...14B}. Switchbacks were suggested to arise as a consequence of solenoidality and incompressibility of the fluctuations \citep{2024PhPl...31c2901M}. It is worth noting that, of the above models, only \citet{2019ApJ...881L...5V} showed the ability to correctly reproduce the statistical intermittency of the fluctuations, necessary for a realistic reproduction of solar wind turbulence.

To capture the multiscale intermittency and non-Gaussian statistics of these fluctuations, modern turbulence models routinely employ multiplicative cascades, most notably the classic binomial p-model \citep{1987PhRvL..59.1424M}. However, applying the classic p-model via linear additive perturbations in $\mathbb{R}^3$ exhibits severe physical and mathematical limitations. First, as demonstrated above, the Euclidean addition of an intermittent cascade inevitably introduces high-amplitude, non-physical compressive variations that are strictly incompatible with the phenomenology of pure Alfv\'enic turbulence. Second, this standard approach conflates variations in vector magnitude with variations in direction. Because the simulated magnitude artificially fluctuates, it completely precludes the rigorous definition of a Waiting Time Distribution (WTD) for pure rotational bursts. When a signal threshold is applied in $\mathbb{R}^{3}$, the identification of scale-independent angular reorientations is hopelessly contaminated by spurious compressive noise, making it mathematically impossible to isolate and study purely directional intermittency. Third, unconstrained flat-space synthesis fails to map the exact curved geometry of the Special Orthogonal Group $SO(3)$, breaking the geometric invariants essential for computing higher-order multipoint structure functions and assessing true spatial correlations in a rotating frame. Indeed, standard Euclidean approaches fundamentally overlook the rich differential geometry underlying ideal fluid dynamics, a connection first established by \citet{1966AnIF...16..319A} and subsequently expanded through the formalization of infinite-dimensional Lie groups and diffeomorphism algebras in continuum mechanics \citep{1970AnMat..92..102E,1983PhyD....7..305M,1998AdMat.137....1H}.

To definitively resolve these topological anomalies without resorting to the computationally expensive numerical integration of full nonlinear magnetohydrodynamic partial differential equations, an advanced rotational p-model is here introduced. The primary purpose of this paper is to formulate a novel, mathematically exact stochastic operator defined on the differentiable manifold of $SO(3)$, capable of generating highly realistic multifractal Alfv\'enic turbulence while rigorously conserving the magnetic field modulus at all scales. By doing so, this approach grounds synthetic turbulence modeling within the robust framework of stochastic geometric mechanics \citep{2014sded.book.....I,2002sam..book.....H} and Lie-group geometric numerical integration \citep{2000AcNum...9..215I,2002SJNA...40.1583M,2006ObeRp...3..805H,2008RpMP...61...65L}, ensuring that the discrete evolution of the stochastic process does not artificially drift from the underlying curved topology.

The paper is structured as follows. \S{} \ref{sec:methodology} introduces the mathematical formulation of the rotational p-model, bridging the discrete multifractal singular measure to the $\mathfrak{so}(3)$ Lie algebra, and detailing the Stratonovich integration of the stochastic differential equations on the $S^{2}$ spherical manifold. \S{} \ref{sec:numerical_experiments} presents the numerical experiments, providing a comprehensive multiscale topological validation of the generated turbulence, including the extraction of multifractal scaling exponents and purely directional waiting time statistics. Finally, \S{} \ref{sec:conclusions} outlines the conclusions and discusses the broader implications for nonlinear heliospheric modeling.

\section{Mathematical methodology}
\label{sec:methodology}
To systematically resolve the topological anomalies inherent in classical Euclidean synthetic models, a novel, geometrically constrained stochastic framework is here presented. The core objective of this methodology is to synthesize a fully developed, intermittent turbulent cascade that acts exclusively as a rotational operator on the magnetic field vector, thereby rigorously preserving the strict kinetic incompressibility ($\nabla\cdot\delta\mathbf{V}=0$) and the exact magnetic modulus conservation ($\delta|\mathbf{B}|=0$) demanded by ideal magnetohydrodynamics. The proposed mathematical architecture is developed through a sequential mapping: first, a non-Gaussian multifractal singular measure is generated to model the localized intermittent dissipation of turbulent energy; second, an innovative spectral filtering technique is applied in the frequency domain to impose the proper fractional scaling of the macroscopic physical observables; finally, the resulting stochastic fields are mapped into the Lie algebra $\mathfrak{so}(3)$ to drive a Stratonovich stochastic differential equation constrained strictly on the spherical manifold $S^{2}$.

\subsection{Classical p-model and multifractal measure}
\label{subsec:p_model}
The synthesis of the stochastic angular vorticity $\boldsymbol{\omega}(t)$ fundamentally relies on the formalism of the multiplicative cascade measure (the so-called p-model) originally introduced by \citet{1987PhRvL..59.1424M} to describe the highly inhomogeneous, intermittent nature of turbulent energy dissipation. Let $\epsilon$ be a strictly positive probability measure representing the turbulent energy dissipation rate, initially defined homogeneously over a macroscopic temporal domain of characteristic integral scale $T$.

The multifractal manifold is generated through an iterative dyadic partition of this domain. This process yields a cascade of discrete scales $l_{n}=T/2^{n}$, where $n$ denotes the generation step of the turbulent cascade. At each arbitrary $n$-th step, the temporal domain is subdivided into $N=2^{n}$ disjoint sub-intervals. The dissipation measure integrated over the $i$-th sub-interval of size $l_{n}$ is denoted as $\mu_{i}(l_{n})$. The anomalous scaling behavior of the turbulence is mathematically captured by the $q$-th order partition function of this measure, which is postulated to obey the following power-law scaling:

\begin{equation}
	Z_{q}(l_{n})=\sum_{i=1}^{2^{n}}\mu_{i}(l_{n})^{q}\sim\left(\frac{l_{n}}{T}\right)^{\tau(q)},
\end{equation}

\noindent where the generalized scaling exponent $\tau(q)$ is a nonlinear convex function of the moment order $q$, signifying the departure from strict scale invariance (monofractality). For the discrete binomial p-model employed in this framework, $\tau(q)$ is analytically determined by the spatial intermittency parameter $p\in[0.5,1]$ as:

\begin{equation}
	\tau(q)=-\log_{2}[p^{q}+(1-p)^{q}].
\end{equation}

Operatively, the specific fraction of turbulent energy assigned to a given segment at the $n$-th scale is constructed through a strictly multiplicative stochastic process. Let $M_{j}$ be independent and identically distributed random multipliers taking the discrete values $2p$ and $2(1-p)$ with equal probability. The localized energy measure $\epsilon_{n}$ at the final, microscopic cascade step $n$ is obtained by the cumulative product of these multipliers along the specific cascade branch:

\begin{equation}
	\epsilon_{n}=\prod_{j=1}^{n}M_{j}.
\end{equation}

The profound multifractal nature of this cascade is geometrically and mathematically described by the singularity spectrum $f(\alpha)$. This spectrum quantifies the fractal dimension of the highly singular subset of intervals characterized by a local Lipschitz-H\"older exponent $\alpha$. The continuous spectrum is derived via the exact Legendre transform of the mass exponent $\tau(q)$:

\begin{equation}
	f(\alpha)=\inf_{q\in\mathbb{R}}\{q\alpha-\tau(q)\}.
\end{equation}

To translate this discrete topological measure into a continuous-time stochastic process suitable for Eulerian observational grids, the $2^{n}$ final discrete values of the cascade $\epsilon_{n}$ are systematically mapped onto a uniformly discretized temporal grid $t_{k}=k\Delta t$, where the sampling resolution is defined by the ultimate cascade scale $\Delta t=T/2^{n}$. This mapping successfully generates a highly intermittent, local energy density time series $\epsilon(t)$.

Finally, the raw spatial components of the angular fluctuations $y_{i}(t)$ (for the Cartesian indices $i\in\{x,y,z\}$) are mathematically modeled as a fractional Brownian process dynamically modulated by this multifractal background structure:

\begin{equation}
	y_{i}(t)=g_{i}(t)\sqrt{\epsilon(t)},
\end{equation}

\noindent where $g_{i}(t)\sim\mathcal{N}(0,1)$ is a normally distributed, strictly uncorrelated Gaussian white noise acting as the underlying stochastic carrier signal. This formulation guarantees that the local variance of the raw rotational fluctuations strictly inherits the non-Gaussian, multifractal topological properties dictated by the underlying p-model cascade.

\subsection{Spectral filtering and covariance tensor}
\label{subsec:spectral_filtering}
The raw intermittent signal $y_{i}(t)$ generated by the multiplicative cascade is fundamentally driven by a Gaussian white noise carrier; consequently, its theoretical power spectral density remains inherently flat and scale-independent. To overcome this topological limitation and accurately impose the correct macroscopic magnetohydrodynamic phenomenology onto the microscopic angular generators, a novel spectral coloring technique is introduced herein. This innovative approach applies a fractional integral operator directly within the frequency domain, explicitly coupling the non-Gaussian spatial intermittency of the p-model with the exact scale-dependent spectral distributions observed in astrophysical space plasmas.

Let $\mathfrak{F}$ denote the continuous Fourier transform operator, such that the mapping into the frequency domain is defined as $\hat{y}_{i}(f)=\mathfrak{F}\{y_{i}(t)\}$. The target spectral energy density of the fully developed turbulence, denoted as $E(f)$, is postulated to follow a distinct broken power-law distribution, mathematically reflecting the multiscale dynamics typical of solar wind measurements:

\begin{equation}
	E(f)\propto
	\begin{cases}
		f^{-1} & \text{if }f< f_{break}\quad\text{(Injection/energy-containing range)}\\
		f^{-5/3} & \text{if }f\ge f_{break}\quad\text{(Magnetohydrodynamic inertial range)}
	\end{cases}
\end{equation}

\noindent where $f$ represents the spacecraft-frame frequency, and $f_{break}$ is the critical spectral break frequency mathematically marking the phenomenological onset of the turbulent cascade.

The spectral shaping is executed by deterministically weighting the complex Fourier amplitudes of the raw intermittent signal with the square root of the prescribed target energy density. The filtered angular components in the frequency domain, denoted as $\hat{\omega}_{i}(f)$, are formally defined as:

\begin{equation}
	\hat{\omega}_{i}(f)=\hat{y}_{i}(f)\sqrt{E(f)}.
\end{equation}

By theoretical construction, the three stochastic angular generators (for the Cartesian indices $i\in\{x,y,z\}$) are assumed to be strictly statistically independent and isotropic within the basis of the Lie algebra space. It is worth noting that while the underlying generators are isotropically defined in this abstract mathematical space, the physical anisotropy typical of Alfv\'enic fluctuations (where perturbations are strictly perpendicular to the mean field) is naturally recovered when these operators act on the physical magnetic field vector background $\mathbf{B}_{0}$, which breaks the spatial symmetry. According to the Wiener-Khinchin theorem for wide-sense stationary stochastic processes, the spectral covariance tensor $\mathfrak{P}_{ij}(f)$ of the filtered signal strictly obeys the following relation:

\begin{equation}
	\langle\hat{\omega}_{i}(f)\hat{\omega}_{j}^{*}(f')\rangle=\delta(f-f')\mathfrak{P}_{ij}(f)\quad\text{with}\quad\mathfrak{P}_{ij}(f)=E(f)\delta_{ij},
\end{equation}

\noindent where $\langle\cdot\rangle$ denotes the temporal ensemble average, $\delta(\cdot)$ represents the Dirac delta distribution, and $\delta_{ij}$ is the Kronecker delta, which mathematically ensures absolute zero cross-correlation between the different spatial axes of rotation.

Finally, to map the rotational dynamics back into the physical time domain, the continuous-time pseudo-vector field $\boldsymbol{\omega}(t)$ is retrieved via the application of the inverse Fourier transform operator ($\mathfrak{F}^{-1}$). To analytically guarantee that the synthesized turbulence possesses a specific, user-defined macroscopic total angular variance $\sigma_{0}^{2}$, the real part of the inverse transformation is rigorously normalized by its $L^{2}$-norm standard deviation:

\begin{equation}
	\omega_{i}(t)=\sigma_{0}\frac{\Re[\mathfrak{F}^{-1}\{\hat{\omega}_{i}(f)\}]}{|\Re[\mathfrak{F}^{-1}\{\hat{\omega}_{i}(f)\}]|_{L^{2}}}.
\end{equation}

This operation yields the exact, energy-preserving components of the instantaneous angular velocity perturbation, ready to be mapped into the algebraic space of rotations.

\subsection{The Lie algebra $\mathfrak{so}(3)$ and generator basis}
\label{subsec:lie_algebra}
The three independent stochastic time series $\omega_{x}(t)$, $\omega_{y}(t)$, and $\omega_{z}(t)$ synthesized in the preceding filtering step represent the exact spatial components of the instantaneous angular velocity perturbation in radians. In the standard Euclidean framework, these components form a time-dependent stochastic pseudo-vector in the flat space $\mathbb{R}^{3}$:

\begin{equation}
	\boldsymbol{\omega}(t)=[\omega_{x}(t),\omega_{y}(t),\omega_{z}(t)]^{T}.
\end{equation}

To mathematically guarantee the absolute topological conservation of the magnetic field modulus (a strict requirement for modeling purely Alfv\'enic, non-compressive turbulence), this translational perturbation cannot be applied additively. Instead, it must be mapped into the algebraic space of rotations, specifically the Lie algebra $\mathfrak{so}(3)$. The space $\mathfrak{so}(3)$ is formally defined as the vector space of all $3\times3$ real skew-symmetric matrices (i.e., matrices satisfying the condition $A^{T}=-A$).

There exists a canonical vector-space isomorphism, frequently denoted in differential geometry as the hat map operator, namely $\hat{}:\mathbb{R}^{3}\rightarrow\mathfrak{so}(3)$, which bijectively maps any $3$-dimensional Euclidean vector to a unique skew-symmetric matrix. To formalize this mapping within the numerical framework, a standard basis for the $\mathfrak{so}(3)$ algebra is introduced, consisting of three infinitesimal generator matrices $J_{1}$, $J_{2}$, and $J_{3}$. The matrix elements of the $i$-th generator are analytically defined using the totally antisymmetric Levi-Civita symbol $\epsilon_{ijk}$:

\begin{equation}
	(J_{i})_{jk}=-\epsilon_{ijk},
\end{equation}

\noindent where the indices $i,j,k\in\{1,2,3\}$ correspond respectively to the Cartesian spatial axes $x,y,z$.

By employing the Einstein summation convention over the repeated index $i$, the stochastic pseudo-vector $\boldsymbol{\omega}(t)$ is rigorously projected into the Lie algebra, yielding the time-dependent perturbative matrix operator $\Omega(t)\in\mathfrak{so}(3)$:

\begin{equation}
	\Omega(t)=\omega_{i}(t)J_{i}=
	\begin{pmatrix}
		0 & -\omega_{z}(t) & \omega_{y}(t)\\
		\omega_{z}(t) & 0 & -\omega_{x}(t)\\
		-\omega_{y}(t) & \omega_{x}(t) & 0
	\end{pmatrix}
\end{equation}

The fundamental algebraic property of this skew-symmetric matrix representation lies in its dynamic action on generic vectors. For any arbitrary column vector $\mathcal{V}\in\mathbb{R}^{3}$, the left-multiplication by the stochastic matrix $\Omega(t)$ is strictly algebraically equivalent to the vector cross product with the angular velocity pseudo-vector $\boldsymbol{\omega}(t)$:

\begin{equation}
	\Omega(t)\mathcal{V}=\boldsymbol{\omega}(t)\times\mathcal{V}.
\end{equation}

This exact mathematical correspondence constitutes the foundational step required to define the nonlinear stochastic differential equation governing the continuous rotational dynamics of the magnetic field vector.

Furthermore, as a formal Lie algebra, $\mathfrak{so}(3)$ is equipped with a non-abelian Lie bracket, defined by the commutator $[\Omega_{A},\Omega_{B}]=\Omega_{A}\Omega_{B}-\Omega_{B}\Omega_{A}$. This bracket inherently captures the fundamental non-commutative nature of macroscopic $3$D rotations, ensuring that the sequence of intermittent turbulent deflections is topologically preserved. The intrinsic geometric structure of this algebraic space is formally characterized by the invariant bilinear Killing form, which acts as a metric tensor:

\begin{equation}
	K(\Omega_{A},\Omega_{B})=-\frac{1}{2}\text{Tr}(\Omega_{A}\Omega_{B}).
\end{equation}

This specific trace form is strictly proportional to the standard Euclidean inner product of their corresponding pseudo-vectors in $\mathbb{R}^{3}$. Consequently, it rigorously bridges the abstract algebraic metric of the curved manifold to the actual physical kinetic energy of the angular turbulent fluctuations driving the system.

\subsection{SDE integration in the Stratonovich sense on the $S^{2}$ manifold}
\label{subsec:sde_integration}
To physically model the temporal evolution of the Alfv\'enic magnetic field driven by the synthesized multifractal turbulence, an infinitesimal stochastic angular increment is formally defined as $d\boldsymbol{\theta}(t)=\boldsymbol{\omega}(t)dt$. The continuous rotation of the macroscopic magnetic field vector $\mathbf{B}(t)$ is consequently governed by a Stochastic Differential Equation (SDE). Geometrically, an infinitesimal rotation in three-dimensional space is expressed via the vector cross product, yielding the governing SDE:

\begin{equation}
	d\mathbf{B}(t)=d\boldsymbol{\theta}(t)\times\mathbf{B}(t).
\end{equation}

This cross-product formulation intrinsically ensures local transversality: the infinitesimal perturbation $d\mathbf{B}(t)$ is perpetually orthogonal to the instantaneous local field vector $\mathbf{B}(t)$, accurately reproducing the shear-driven, non-compressive nature of nonlinear Alfv\'enic fluctuations.

By substituting the vector cross product with the algebraically equivalent left-multiplication by the skew-symmetric matrix operator $\Omega(t)\in\mathfrak{so}(3)$ defined in the previous section, the equation is recast into a strict matrix formulation. However, to mathematically guarantee that the integration of this stochastic process geometrically confines the evolving vector onto the two-dimensional Riemannian spherical manifold $S^{2}$ (which is smoothly embedded in $\mathbb{R}^{3}$), the SDE must be rigorously interpreted in the sense of Fisk-Stratonovich calculus (customarily denoted by the operator $\circ$).

The Stratonovich integration scheme is strictly required in this topological context because, unlike classical It\^o calculus, it uniquely preserves the ordinary rules of differential geometry, specifically, the standard chain rule. The It\^o formulation, while possessing favorable martingale properties, would inadvertently introduce spurious stochastic drift terms (originating from the non-zero second-order variations) that would systematically drive the vector off the spherical surface, thereby violating the physical manifold constraints. The formal Stratonovich SDE is therefore written as:

\begin{equation}
	d\mathbf{B}(t)=\Omega(t)\circ\mathbf{B}(t)dt.
\end{equation}

The topological invariance of the magnetic field modulus, which physically represents the perfect kinetic incompressibility of the Alfv\'enic plasma, can now be analytically proven. Let the squared Euclidean norm of the magnetic field be defined via the inner product $|\mathbf{B}(t)|^{2}=\langle\mathbf{B}(t),\mathbf{B}(t)\rangle=\mathbf{B}(t)^{T}\mathbf{B}(t)$. Applying the Stratonovich differential operator to this scalar quantity yields:

\begin{equation}
	d(|\mathbf{B}(t)|^{2})=2\langle\mathbf{B}(t),\circ d\mathbf{B}(t)\rangle.
\end{equation}

Substituting the governing SDE into this exact differential expression provides the evolution of the norm:

\begin{equation}
	d(|\mathbf{B}(t)|^{2})=2\langle\mathbf{B}(t),\Omega(t)\mathbf{B}(t)\rangle dt.
\end{equation}

By the fundamental theorems of linear algebra, the quadratic form associated with any real skew-symmetric matrix $\Omega(t)$ and any real vector $\mathbf{B}(t)$ is identically zero (i.e., $\mathbf{B}(t)^{T}\Omega(t)\mathbf{B}(t)\equiv0$). Consequently, the differential strictly and unconditionally collapses to:

\begin{equation}
	d(|\mathbf{B}(t)|^{2})=0.
\end{equation}

This rigorous mathematical result demonstrates that the phase orbit of the stochastically perturbed vector perpetually resides on the spherical geodesic of constant radius $R=|\mathbf{B}_{0}(t)|$. The Stratonovich formulation on the Lie algebra inherently prevents any spurious fluctuations of the magnetic magnitude, providing a perfect geometric and topological description of purely Alfv\'enic, incompressible turbulence within the framework of nonlinear dynamical systems.

\subsection{Exponential map and Cayley-Hamilton theorem}
\label{subsec:exponential_map}
The exact analytical resolution of the Stratonovich stochastic differential equation requires projecting the perturbation from the tangent space defined by the Lie algebra $\mathfrak{so}(3)$ onto the continuous Lie group $SO(3)$. This projection is formally achieved via the matrix exponential map, which provides the fundamental mathematical bridge between the local linear perturbations and the global nonlinear spherical manifold.

Let $I$ denote the $3\times3$ identity matrix. The magnitude of the instantaneous stochastic angular velocity vector is defined as the scalar angle $\theta(t)=|\boldsymbol{\omega}(t)|=\sqrt{\omega_{x}^{2}+\omega_{y}^{2}+\omega_{z}^{2}}$, and the corresponding normalized axis of continuous rotation is represented by the unit vector $\mathbf{k}(t)=\boldsymbol{\omega}(t)/\theta(t)$.

The eigenvalue spectrum, denoted by $\lambda$, of the skew-symmetric matrix $\Omega$ is determined by the characteristic equation $\det(\Omega-\lambda I)=0$. Expanding this determinant for the specific stochastic matrix $\Omega$ yields the characteristic polynomial:

\begin{equation}
	\lambda^{3}+\theta^{2}\lambda=0.
\end{equation}

According to the Cayley-Hamilton theorem, every square matrix strictly satisfies its own characteristic equation. Substituting the matrix operator $\Omega$ for the scalar eigenvalue $\lambda$ yields a fundamental matrix recursion relation:

\begin{equation}
	\Omega^{3}+\theta^{2}\Omega=0.
\end{equation}

This deterministic recursive property allows higher-order matrix powers to be systematically and algebraically reduced. Specifically, odd powers identically collapse to $\Omega^{2m+1}=(-1)^m\theta^{2m}\Omega$, and even powers geometrically collapse to $\Omega^{2m+2}=(-1)^{m}\theta^{2m}\Omega^{2}$ for any integer $m\ge0$.

The Lie group rotation operator $\mathcal{R}_{\mathbf{k},\theta}(t)\in SO(3)$ is defined by the infinite Taylor series expansion of the matrix exponential:

\begin{equation}
	\mathcal{R}_{\mathbf{k},\theta}(t)=\exp(\Omega)=\sum_{n=0}^{\infty}\frac{\Omega^{n}}{n!}=I+\Omega+\frac{\Omega^{2}}{2!}+\frac{\Omega^{3}}{3!}+\frac{\Omega^{4}}{4!}+\dots
\end{equation}

By partitioning the infinite summation into the identity matrix, odd powers, and even powers, and subsequently substituting the recursion relation derived above, the series factors into recognized Maclaurin expansions for the sine and cosine trigonometric functions:

\begin{equation}
	\exp(\Omega)=I+\frac{1}{\theta}\left(\theta-\frac{\theta^{3}}{3!}+\frac{\theta^{5}}{5!}-\dots\right)\Omega+\frac{1}{\theta^{2}}\left(\frac{\theta^{2}}{2!}-\frac{\theta^{4}}{4!}+\frac{\theta^{6}}{6!}-\dots\right)\Omega^{2}.
\end{equation}

This infinite series exactly and unconditionally converges to a finite matrix polynomial, universally known in differential geometry as Rodrigues' rotation formula:

\begin{equation}
	\mathcal{R}_{\mathbf{k},\theta}(t)=I+\left(\frac{\sin\theta}{\theta}\right)\Omega+\left(\frac{1-\cos\theta}{\theta^{2}}\right)\Omega^{2}.
\end{equation}

To obtain the final physical macroscopic field, this rigorous $SO(3)$ matrix operator is applied directly to the background magnetic field vector $\mathbf{B}_{0}(t)$. Recalling the established isomorphism $\Omega\mathbf{B}_{0}=\boldsymbol{\omega}\times\mathbf{B}_{0}$ (where $\boldsymbol{\omega}=\theta\mathbf{k}$), and analytically applying the vector triple product expansion for the squared term, $\Omega^{2}\mathbf{B}_{0}=\boldsymbol{\omega}\times(\boldsymbol{\omega}\times\mathbf{B}_{0})=\theta^{2}[\mathbf{k}(\mathbf{k}\cdot\mathbf{B}_{0})-\mathbf{B}_{0}]$, the substitution translates the matrix operator directly into an exact geometric integrator vector form:

\begin{equation}
	\mathbf{B}(t)=\mathbf{B}_{0}(t)\cos\theta(t)+(\mathbf{k}(t)\times\mathbf{B}_{0}(t))\sin\theta(t)+\mathbf{k}(t)(\mathbf{k}(t)\cdot\mathbf{B}_{0}(t))(1-\cos\theta(t)).
\end{equation}

This exact topological rotation analytically guarantees the strict conservation of the local magnetic field modulus at all continuous time steps, entirely obviating the need for computationally dissipative or geometrically unstable iterative numerical integration schemes.

\section{Numerical experiments}
\label{sec:numerical_experiments}
To rigorously validate the theoretical framework and empirically demonstrate the topological preservation capabilities of the Lie-group mapping, a comprehensive set of numerical experiments is performed. The synthetic time series is discretized over a uniformly sampled temporal grid consisting of $N=8192$ points with an arbitrary temporal resolution $\Delta t=1$. The macroscopic background magnetic field is initialized as a purely longitudinal vector $\mathbf{B}_{0}=(10,0,0)$. To simulate a highly intermittent, fully developed Alfv\'enic cascade, the spatial intermittency parameter of the underlying p-model is set to $p=0.8$ \citep{1997NPGeo...4..185H}, while the spectral break marking the transition from the $f^{-1}$ injection range to the $f^{-5/3}$ inertial range is positioned at $f_{break}=10^{-2}$ \citep{2013LRSP...10....2B}. The root-mean-square amplitude of the angular fluctuations is prescribed as $\sigma_{rot}=0.4$ radians. Geometrically, under the constraint of constant magnetic field magnitude characteristic of clean Alfv\'enic fluctuations \citep{2019NCimC..42...16M}, this angular variance corresponds to a normalized fluctuation amplitude of $|\delta\mathbf{B}|/|\mathbf{B}_0|=2\sin(\sigma_{rot}/2)\approx0.4$, which is highly consistent with typical large-amplitude observations in the fast solar wind.

\subsection{Topological invariants, spectral scaling, and scale-dependent flatness}
\label{subsec:flatness}
\begin{figure}[ht]
	\centering
	\includegraphics[width=0.8\textwidth]{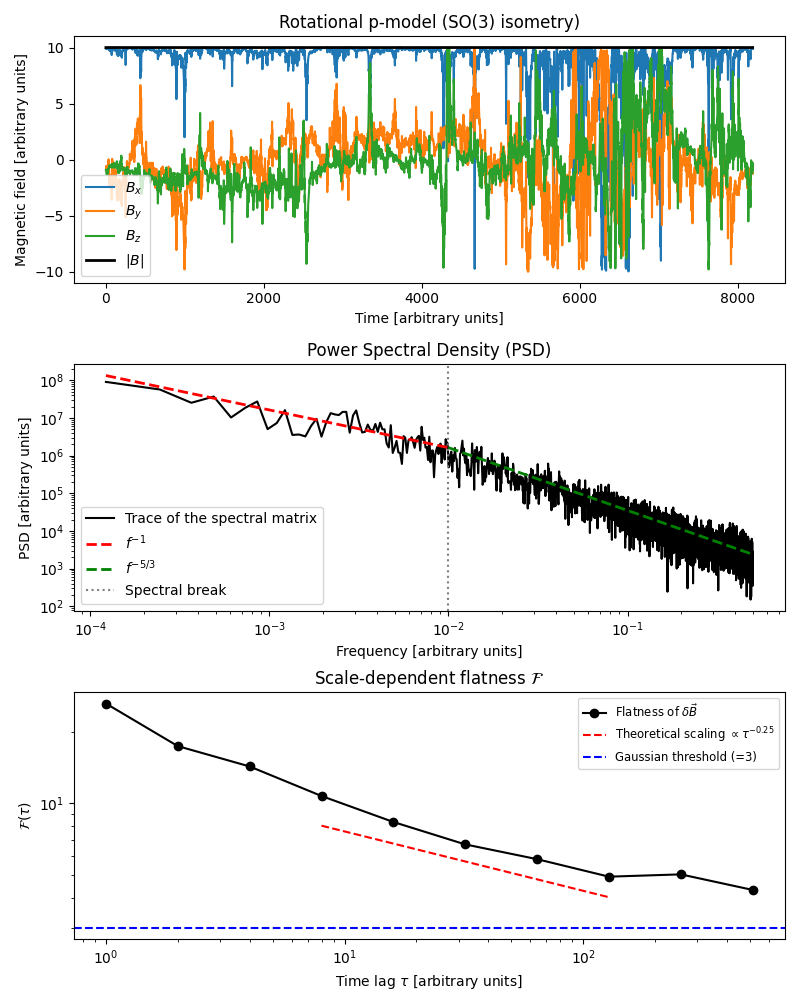}
	\caption{Multiscale topological validation of the $SO(3)$ rotational p-model. \emph{Top panel:} synthetic time series of the macroscopic magnetic field Cartesian components ($B_{x}$,$B_{y}$,$B_{z}$) and the scalar magnitude ($|\mathbf{B}|$). The strictly constant $|\mathbf{B}|$ empirically confirms the rigorous preservation of the geometric isometry and the exact kinetic incompressibility of the simulated plasma. \emph{Middle panel:} trace of the PSD matrix, explicitly highlighting the $f^{-1}$ energy injection range (marked with a red dashed line), the $f^{-5/3}$ Kolmogorov-like inertial range (marked with a green dashed line), and the prescribed spectral break $f_{break}$ (vertical dotted line). \emph{Bottom panel:} scale-dependent flatness $\mathcal{F}(\tau)$ computed from the purely directional vector increments $\delta\mathbf{B}$. The marked divergence from the Gaussian baseline ($\mathcal{F}=3$, horizontal blue dashed line) at small temporal scales ($\tau\to0$) mathematically demonstrates the successful injection of non-Gaussian angular intermittency. The red dashed line illustrates the theoretical scaling prediction $\mathcal{F}(\tau)\propto\tau^{-0.25}$ for the inertial range ($\tau\in[8,128]$), which is analytically derived from the multifractal structure functions of the p-model by assuming an intermittency parameter $p=0.80$ and a Kolmogorov scaling $h=1/3$.}
	\label{fig:multiscale_validation}
\end{figure}

The primary validation of the rotational p-model lies in its capability to couple complex multiscale stochastic fluctuations with strict geometric invariants. Figure \ref{fig:multiscale_validation} (top panel) illustrates the macroscopic temporal profiles of the three Cartesian magnetic field components ($B_{x},B_{y},B_{z}$) alongside the total magnetic field magnitude $|\mathbf{B}|$ synthesized by the $\mathfrak{so}(3)$ Stratonovich SDE integration.

While the individual vector components exhibit large-amplitude, highly irregular continuous deflections characteristic of intermittent turbulence, the scalar magnitude $|\mathbf{B}|$ remains perfectly constant at the prescribed background value ($|\mathbf{B}|=10$) throughout the entire temporal domain. This is qualitatively reminiscent of large-amplitude Alfv\'enic fluctuations (such as switchbacks) observed in the solar wind \citep[see][an references therein]{2026SSRv..222...14B}. This result provides empirical proof that the Lie-group stochastic mapping acts as a pure geometric isometry, definitively preventing the generation of spurious compressive fluctuations that inherently plague classical Euclidean additive models.

The accuracy of the frequency-domain spectral filtering is confirmed by evaluating the total turbulent spectral energy, defined mathematically as the trace of the spectral covariance matrix. Let $\delta B_{i}(t)=B_{i}(t)-\langle B_{i}\rangle$ represent the zero-mean fluctuating part of the $i$-th macroscopic magnetic field Cartesian component. The trace of the Power Spectral Density (PSD) matrix is analytically computed as the sum of the individual squared Fourier amplitudes (auto-spectra):

\begin{equation}
	\text{PSD}(f)=\sum_{i\in\{x,y,z\}}|\mathfrak{F}\{\delta B_{i}(t)\}|^{2}.
\end{equation}

The synthetic spectrum derived from this trace precisely conforms to the theoretical fractional broken power-law envelope (Figure \ref{fig:multiscale_validation}, middle panel). At frequencies below $f_{break}$, the energy distribution correctly scales as $f^{-1}$, representing the injection scales of coronal origin, while the higher-frequency domain exhibits a rigorous $f^{-5/3}$ Kolmogorov-like scaling, confirming the correct macroscopic energy partition of the inertial range.

To statistically quantify the presence of small-scale intermittency, the scale-dependent flatness (or fourth-order standardized moment) $\mathcal{F}(\tau)$ is evaluated. This metric requires the computation of the two-point temporal vector increment at a given scale lag $\tau$, formally defined as:

\begin{equation}
	\Delta\mathbf{B}(t,\tau)=\mathbf{B}(t+\tau)-\mathbf{B}(t).
\end{equation}

The flatness is subsequently computed as the ratio of the fourth-order structure function to the square of the second-order structure function:

\begin{equation}
	\mathcal{F}(\tau)=\frac{\langle|\Delta\mathbf{B}(t,\tau)|^{4}\rangle}{\langle|\Delta\mathbf{B}(t,\tau)|^{2}\rangle^{2}},
\end{equation}

\noindent where $\langle\cdot\rangle$ denotes the temporal ensemble average.

From a purely mathematical standpoint, the Euclidean vector increment $\Delta\mathbf{B}(t,\tau)$ natively incorporates arbitrary variations in both direction and magnitude. However, a fundamental topological advantage emerges within the rotational p-model framework: because the continuous model applies a strict $SO(3)$ isometry, the local magnetic field magnitude is an absolute geometric invariant ($|\mathbf{B}(t)|=|\mathbf{B}(t+\tau)|=\text{const}$). Consequently, the geometric displacement $|\Delta\mathbf{B}(t,\tau)|$ depends exclusively on the temporal angular deflection between the two states. This implies that the resulting structure functions and the scale-dependent flatness isolate and quantify exclusively the purely directional intermittency of the dynamical system, remaining completely decoupled from any compressive scalar noise.

Figure \ref{fig:multiscale_validation} (bottom panel) displays the computed flatness $\mathcal{F}(\tau)$ across various characteristic time lags. At large scales ($\tau\gg1/f_{break}$), the flatness converges strictly to the Gaussian baseline ($\mathcal{F}=3$), indicating that large-scale field reorientations are statistically independent and normally distributed. Conversely, as the time scale progressively decreases deep into the inertial range ($\tau\to0$), $\mathcal{F}(\tau)$ exhibits a sharp, monotonic divergence well above the Gaussian threshold. In the inertial regime ($\tau\in[8,128]$), this scale-dependent departure is governed by the underlying multiplicative cascade and closely follows the theoretical scaling prediction $\mathcal{F}(\tau)\propto\tau^{\kappa}$. The scaling exponent $\kappa$ can be analytically derived from the structure function exponents as $\kappa=\zeta_{4}-2\zeta_{2}$. Using the standard framework of the p-model, this relation expands to $\kappa=2\log_{2}\left[p^{2h}+(1-p)^{2h}\right]-\log_{2}\left[p^{4h}+(1-p)^{4h}\right]-1$. By substituting the characteristic Kolmogorov Hurst exponent $h=1/3$ for the forced inertial spectrum and the herein imposed intermittency parameter $p=0.80$, a theoretical scaling exponent of $\kappa\approx-0.25$ is obtained. The alignment of the computed flatness with this theoretical reference curve, mathematically proves that the synthetic directional fluctuations successfully inherit the non-Gaussian, singular topology of the multiplicative cascade, correctly manifesting the intermittent angular bursts that characterize true heliospheric Alfv\'enic turbulence.

\subsection{Probability Density Functions of projected field increments}
\label{subsec:pdfs}
A fundamental statistical signature of fully developed turbulence is the scale-dependent deformation of the Probability Density Functions (PDFs) of the field increments. In classical magnetohydrodynamics, intermittency manifests as a progressive departure from Gaussian statistics at large spatial or temporal scales toward strongly leptokurtic (heavy-tailed) distributions at smaller scales within the inertial range. To rigorously evaluate this topological transition within the $SO(3)$ mapped fields, the probability distributions of the projected vector increments are computed.

While the exact vector increment $\Delta\mathbf{B}(t,\tau)$ captures the full three-dimensional geometric displacement, the purely transversal nature of Alfv\'enic fluctuations is most accurately analyzed by projecting this increment onto a manifold orthogonal to the macroscopic background magnetic field. Given the strictly longitudinal background field configuration $\mathbf{B}_{0}=(B_{0},0,0)$, a transverse unit projection vector is formally defined as $\mathbf{e}_{\text{proj}}=(0,1/\sqrt{2},1/\sqrt{2})$. The scalar projected increment is then mathematically computed via the inner product:

\begin{equation}
	\delta B_{\text{proj}}(t,\tau)=\Delta\mathbf{B}(t,\tau)\cdot\mathbf{e}_{\text{proj}}.
\end{equation}

To systematically enable a rigorous comparative analysis of the probability shapes across vastly distinct temporal scales, these projected increments are statistically standardized to zero mean and unit variance:

\begin{equation}
\delta\tilde{B}_{\text{proj}}(t,\tau)=\frac{\delta B_{\text{proj}}(t,\tau)-\langle\delta B_{\text{proj}}(t,\tau)\rangle}{\sigma_{\text{proj}}(\tau)},
\end{equation}

\noindent where $\sigma_{\text{proj}}(\tau)$ denotes the exact standard deviation of the projected increment ensemble at the specific scale lag $\tau$.

\begin{figure}[ht]
	\centering
	\includegraphics[width=0.8\textwidth]{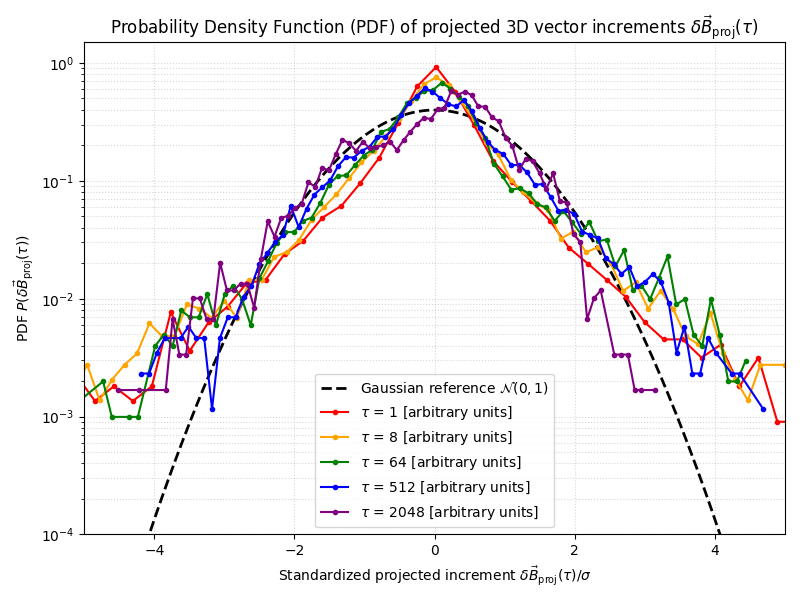}
	\caption{PDFs of the standardized projected increments $\delta\tilde{B}_{\text{proj}}(\tau)$ for various temporal lags $\tau$. The dashed black curve represents the standard Gaussian reference $\mathcal{N}(0,1)$. The progressive evolution from a nearly Gaussian profile at large scales ($\tau=2048$) to highly leptokurtic shapes with pronounced heavy tails at small scales ($\tau\to1$) rigorously demonstrates the scale-dependent non-Gaussian intermittency generated by the continuous $SO(3)$ rotational cascade.}
	\label{fig:pdf_increments}
\end{figure}

Figure \ref{fig:pdf_increments} displays the computed PDFs of the standardized projected increments for a sequence of temporal lags $\tau$ ($=1,\,8,\,64,\,512,\text{ and }2028$). The standard Gaussian distribution $\mathcal{N}(0,1)$ is intrinsically provided as a baseline theoretical reference. At large scales ($\tau=2048$), the probability distribution closely converges to the Gaussian parabola, mathematically indicating that widely separated angular deflections effectively represent a sum of independent, completely uncorrelated random rotations. However, as the temporal lag is systematically reduced deep into the inertial range ($\tau\in\{1,8,64\}$), the core of the distribution sharpens considerably, and pronounced heavy tails dynamically emerge. This continuous, scale-dependent topological deformation provides definitive evidence that the $SO(3)$ rotational p-model accurately preserves and dynamically reproduces the intermittent, non-Gaussian probability statistics strictly characteristic of localized coherent structures in astrophysical Alfv\'enic turbulence.

\subsection{Multifractal spectrum and anomalous scaling exponents}
\label{subsec:multifractal_spectrum}
To comprehensively characterize the scale-invariance and the turbulent energy transfer mechanisms, the multiscale statistical properties of the synthetic field are analyzed via the generalized structure functions. Unlike the projected increments utilized for the probability density functions, the structure functions are computed using the full magnitude of the three-dimensional vector increment, defined as $|\Delta\mathbf{B}(t,\tau)|=|\mathbf{B}(t+\tau)-\mathbf{B}(t)|$. As discussed above, because the $SO(3)$ rotational p-model strictly conserves the magnetic field norm, this scalar magnitude physically represents the exact chordal distance between two states on the $S^{2}$ manifold, thereby serving as a pure, dynamically rigorous metric for the instantaneous angular deflection.

The $q$-th order structure function $S_{q}(\tau)$ is formally defined as the temporal ensemble average of the $q$-th power of the vector increment magnitude:

\begin{equation}
	S_{q}(\tau)=\langle|\Delta\mathbf{B}(t,\tau)|^{q}\rangle.
\end{equation}

In the classical theory of fully developed turbulence, structure functions evaluated within the inertial range (i.e., at scales considerably smaller than the correlation length but larger than the dissipation scale) exhibit a power-law dependency on the temporal lag $\tau$:

\begin{equation}
	S_{q}(\tau)\propto\tau^{\zeta(q)},
\end{equation}

\noindent where $\zeta(q)$ represents the anomalous scaling exponent of order $q$. For a globally scale-invariant, non-intermittent process characterized by a $f^{-5/3}$ energy spectrum, the Kolmogorov 1941 \citep[K41;][]{1941DoSSR..30..301K} phenomenological theory predicts a strictly linear, monofractal scaling strictly dictated by $\zeta(q)=q/3$. Conversely, the presence of localized intermittent structures intrinsically breaks this global self-similarity, resulting in a strictly nonlinear, convex function $\zeta(q)$.

To empirically extract the scaling exponents from the synthesized data, the structure functions are computed for moment orders ranging from $q=1$ to $q=6$ over temporal scales strictly confined within the inertial range ($\tau\le 1/f_{break}\implies\tau\le 64$). The exponents $\zeta(q)$ are subsequently determined by computing the linear least-squares fit of $\log_{10}S_{q}(\tau)$ versus $\log_{10}\tau$.

\begin{figure}[ht]
	\centering
	\includegraphics[width=0.8\textwidth]{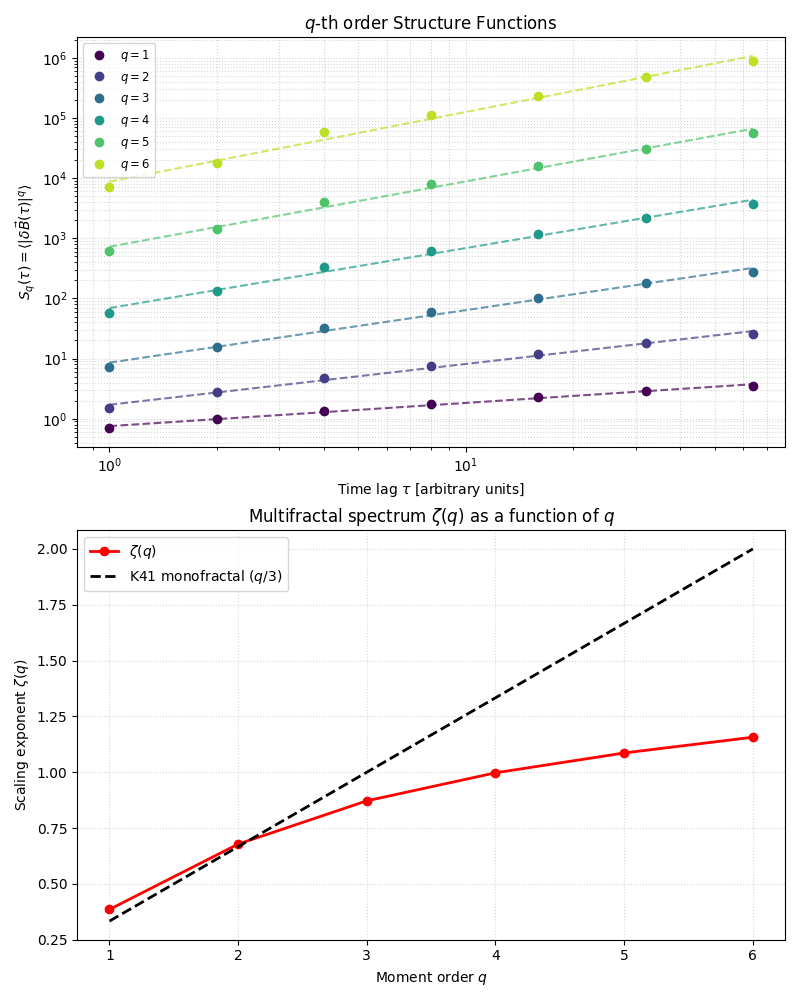}
	\caption{Multifractal scaling analysis of the purely directional intermittent fluctuations. \emph{Top panel:} $q$-th order structure functions $S_{q}(\tau)$ computed for the magnitude of the $3$D vector increments, displayed in log-log scale for moment orders $q\in\{1,\dots,6\}$ across the inertial range. The dashed lines represent the corresponding linear power-law fits. \emph{Bottom panel:} extracted anomalous scaling exponents $\zeta(q)$ as a strictly convex function of the moment order $q$ (solid red line with markers). The prominent deviation from the K41 monofractal theoretical prediction ($\zeta(q)=q/3$, dashed black line) rigorously confirms the multifractal topology of the generated rotational cascade.}
	\label{fig:multifractal_spectrum}
\end{figure}

Figure \ref{fig:multifractal_spectrum} (top panel) displays the scaling of the computed structure functions, confirming robust power-law behaviors across the selected inertial sub-range. The extracted multifractal spectrum $\zeta(q)$ is presented in Figure \ref{fig:multifractal_spectrum} (bottom panel) and directly compared against the linear K41 monofractal theoretical prediction. The evaluated scaling exponents exhibit a pronounced convex deviation from the linear theoretical expectation as the moment order $q$ increases. This fundamental nonlinear mathematical signature rigorously proves that the topological mapping of the multiplicative cascade onto the $SO(3)$ algebra accurately injects and preserves the anomalous scaling laws characterizing the multifractal nature of space plasmas.

\subsection{Waiting time statistics of directional intermittent bursts}
\label{subsec:waiting_times}
A fundamental limitation of classical Euclidean additive models is their inherent inability to accurately evaluate the Waiting Time Distribution (WTD) of intermittent magnetic bursts. In such unconstrained spatial models, the artificial fluctuations of the magnetic field magnitude inextricably pollute the thresholding process, rendering it mathematically impossible to isolate pure orientational intermittency. Conversely, because the proposed $SO(3)$ mapped SDE strictly conserves the magnetic field norm at all scales, the temporal displacement entirely represents an exact angular deflection, providing a robust, unambiguous topological metric for the burst statistics.

To systematically evaluate the WTD, the analysis focuses on the temporal recurrence of extreme localized gradients. The fundamental increment is computed at the minimum available temporal resolution (the integration step $\tau=\Delta t$):

\begin{equation}
	\Delta\mathbf{B}(t,\Delta t)=\mathbf{B}(t+\Delta t)-\mathbf{B}(t).
\end{equation}

Consistent with the probability density analysis, this increment is orthogonally projected onto the transverse manifold via the unit vector $\mathbf{e}_{\text{proj}}$, yielding the scalar projected derivative $\delta B_{\text{proj}}(t)$.

An intermittent directional burst is formally defined to occur at a discrete temporal index $t_{k}$ when the absolute magnitude of this localized gradient strictly exceeds a predefined high-amplitude threshold. To specifically isolate the highly singular, non-Gaussian structures characterizing the multifractal cascade, the intermittency threshold is analytically set to two standard deviations ($2\sigma_{\text{proj}}$) of the projected fluctuation ensemble:

\begin{equation}
	|\delta B_{\text{proj}}(t_{k})|>2\sigma_{\text{proj}}.
\end{equation}

The waiting time $\Delta T_{i}$ separating two consecutive intermittent directional bursts is subsequently defined as the temporal difference between their respective occurrences:

\begin{equation}
	\Delta T_{i}=t_{k+1}-t_{k}.
\end{equation}

In standard memoryless dynamical systems, the distribution of waiting times strictly obeys Poissonian statistics, yielding an exponential probability density function $\mathcal{P}(\Delta T)=\lambda\exp(-\lambda\Delta T)$, where $\lambda$ represents the mean occurrence rate. However, fully developed turbulence is characterized by long-range temporal correlations and clustered intermittent structures.

\begin{figure}[ht]
	\centering
	\includegraphics[width=0.8\textwidth]{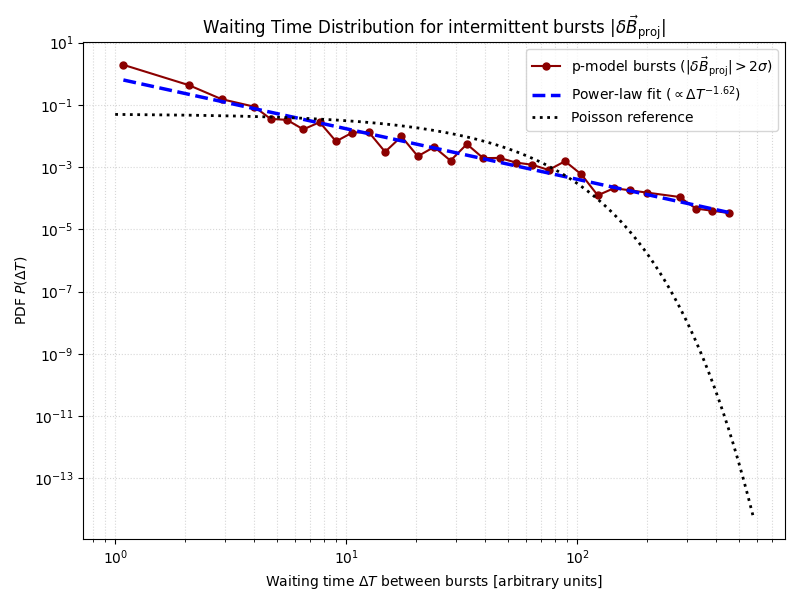}
	\caption{WTD of the intermittent directional bursts, evaluated at the minimum temporal scale ($\tau=\Delta t$). The bursts are systematically extracted by applying an amplitude threshold ($|\delta B_{\text{proj}}(t_{k})|>2\sigma_{\text{proj}}$). The empirical probability density (solid purple line with markers) strongly diverges from the memoryless exponential Poisson distribution (dotted black line), strictly conforming to a power-law asymptotic scaling $\mathcal{P}(\Delta T)\propto\Delta T^{-1.62}$ (dashed blue line). This robust power-law signature mathematically validates the long-range temporal memory and topological clustering of the continuous $SO(3)$ angular deflections.}
	\label{fig:waiting_time}
\end{figure}

Figure \ref{fig:waiting_time} illustrates the probability density function of the computed waiting times $\mathcal{P}(\Delta T)$ in a logarithmic spatial representation. The statistical distribution exhibits a massive departure from the memoryless Poissonian theoretical reference. Instead, the recurrence of the topological bursts rigorously follows a continuous power-law decay:

\begin{equation}
	\mathcal{P}(\Delta T)\propto\Delta T^{-\gamma},
\end{equation}

\noindent where the spectral index is empirically evaluated via linear least-squares fitting as $\gamma\approx1.62$. The emergence of this strict power-law scaling in the temporal domain mathematically demonstrates that the generated Alfv\'enic turbulence possesses deep, long-range memory, successfully translating the geometric clustering of the multifractal p-model into physically accurate, scale-free rotational persistence. Crucially, this scale-free behavior in the WTD is not a generic byproduct of the model's setup, but is deeply coupled with the conservation properties of the generated field. If geometric or physical constraints (such as $|\mathbf{B}|\approx\text{const.}$) are poorly handled, spurious fluctuations can artificially shift clean, low-amplitude rotational structures above the intermittency detection threshold, distorting the waiting time statistics especially at short-to-intermediate timescales.

\subsection{Intermittent angular deflection dynamics}
\label{subsec:angular_deflection}
To visually and mathematically capture the temporal clustering of intermittent structures dynamically evolving on the spherical manifold, the instantaneous angular deflection of the magnetic field vector is directly analyzed. While the structure functions evaluated in \S{} \ref{subsec:multifractal_spectrum} quantify the statistical properties of the increments over varying temporal lags, it is instructive to resolve the exact contiguous orientational jumps occurring at the minimum integration step $\Delta t$.

The instantaneous angular deflection $\Delta\theta(t)$ between successive discrete temporal states $\mathbf{B}(t)$ and $\mathbf{B}(t+\Delta t)$ is computed via the standard Euclidean dot product. Because the Lie-group Stratonovich integration rigorously and unconditionally preserves the macroscopic magnetic field norm $|\mathbf{B}|$ at all times, the exact instantaneous rotational angle is algebraically derived without spurious scalar contributions as:

\begin{equation}
	\Delta\theta(t)=\arccos\left(\frac{\mathbf{B}(t)\cdot\mathbf{B}(t+\Delta t)}{|\mathbf{B}|^2}\right).
\end{equation}

To systematically isolate extreme rotational bursts from the background continuous stochastic drift, a stringent intermittency threshold is established. This threshold is mathematically defined as two standard deviations above the temporal mean of the angular deflection continuous series:

\begin{equation}
	\Delta\theta_{\text{threshold}}=\langle\Delta\theta\rangle+2\sigma_{\Delta\theta}.
\end{equation}

\begin{figure}[ht]
	\centering
	\includegraphics[width=0.8\textwidth]{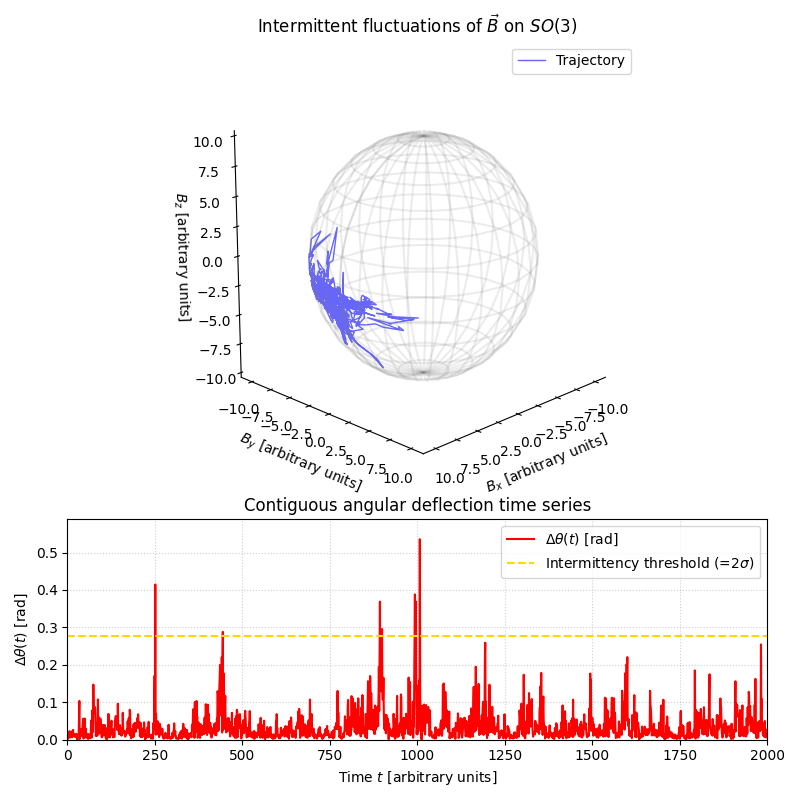}
	\caption{Continuous topological evolution and intermittent angular deflection dynamics. \emph{Top panel:} three-dimensional trajectory of the magnetic field vector exactly constrained on the $S^{2}$ spherical manifold, illustrating the strict conservation of the kinetic incompressibility invariant. \emph{Bottom panel:} time series of the instantaneous contiguous angular deflection $\Delta\theta(t)$ evaluated at the minimum temporal resolution $\Delta t$. The horizontal dashed yellow line dictates the $2\sigma_{\Delta\theta}$ intermittency threshold. The highly clustered, severe bursts traversing this limit dynamically demonstrate the underlying multifractal topology of the cascade \textit{(a full dynamic $3$D animation of this temporal evolution is available in the online supplementary material of the journal)}.}
	\label{fig:angular_dynamics}
\end{figure}

Figure \ref{fig:angular_dynamics} presents the dynamic topological evolution of the synthesized magnetic field. The top panel projects the resulting three-dimensional trajectory of the vector tip directly onto the $S^{2}$ spherical manifold, illustrating the geometric confinement mathematically enforced by the $\mathfrak{so}(3)$ mapped operator. The bottom panel displays the corresponding strictly positive time series of the instantaneous contiguous angular deflections $\Delta\theta(t)$. The temporal profile exhibits a highly inhomogeneous distribution of rotational energy. Long periods of relatively smooth, low-amplitude topological drift are abruptly interrupted by clusters of sharp, violent spikes that massively exceed the $2\sigma_{\Delta\theta}$ threshold. This definitive visual and analytical representation corroborates the multiscale statistical findings, confirming that the synthesized $SO(3)$ Alfv\'enic cascade accurately replicates the severe temporal clustering and spatial inhomogeneity inherently characteristic of extreme nonlinear space plasmas.

\subsection{Probability density and ergodic exploration of the $S^{2}$ manifold}
\label{subsec:ergodic_exploration}
To synthesize the global macroscopic behavior of the rotational cascade, the spatial probability distribution of the stochastically perturbed magnetic field is evaluated directly on the surface of the invariant manifold. While the previous sections analyzed the temporal structure of the increments, characterizing the ergodic exploration of the phase space provides critical insight into the long-term spatial clustering of the localized turbulent states.

The continuous three-dimensional Cartesian time series $\mathbf{B}(t)=(B_{x}(t),B_{y}(t),B_{z}(t))$ is analytically mapped into standard spherical coordinates. Leveraging the mathematically proven invariant radius $R=|\mathbf{B}|$, the instantaneous state of the system is uniquely defined by the azimuthal angle $\varphi(t)$ and the polar angle $\vartheta(t)$:

\begin{equation}
	\varphi(t)=\arctan\left(\frac{B_{y}(t)}{B_{x}(t)}\right)\quad\in[-\pi,\pi]
\end{equation}

\begin{equation}
	\vartheta(t)=\arccos\left(\frac{B_{z}(t)}{R}\right)\quad\in[0,\pi].
\end{equation}

The empirical probability density function $\mathcal{P}(\varphi,\vartheta)$ representing the relative residency time of the vector tip is subsequently constructed. The spherical domain is uniformly discretized, and the bivariate occurrence density is computed by accumulating the instantaneous spatial configurations. To visualize the relative occupancy across the manifold, the raw density is normalized by its absolute maximum, yielding a dimensionless relative density metric bounded within the interval $[0,1]$. This bounded probability map is finally projected back onto the $3$D surface defined by $X_{surf}=R\cos\varphi\sin\vartheta$, $Y_{surf}=R\sin\varphi\sin\vartheta$, and $Z_{surf}=R\cos\vartheta$.

\begin{figure}[ht]
	\centering
	\includegraphics[width=0.8\textwidth]{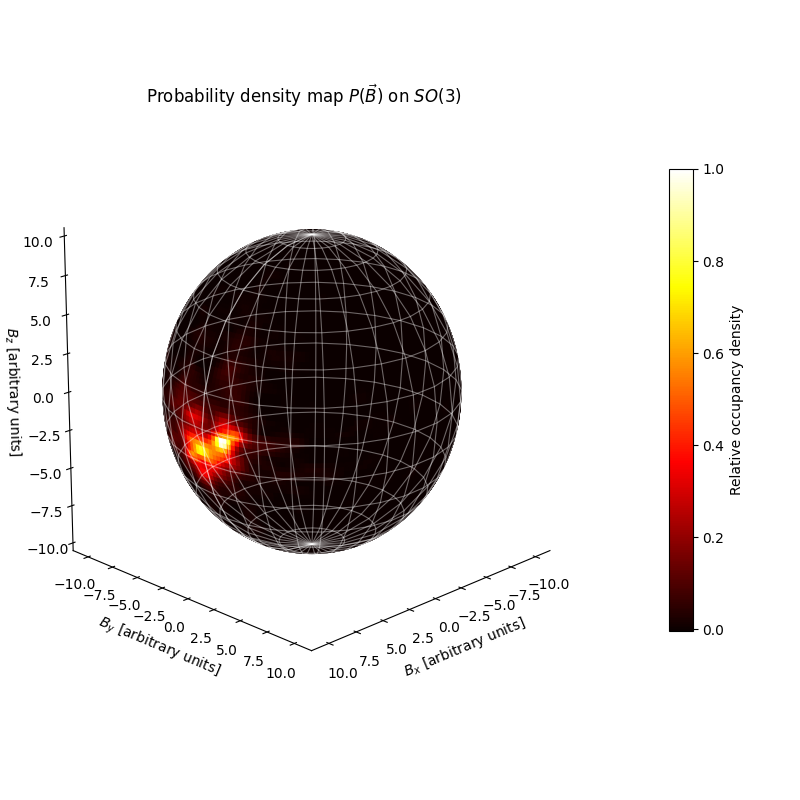}
	\caption{Three-dimensional probability density map $P(\mathbf{B})$ of the dynamically evolving magnetic field on the $S^{2}$ continuous manifold. The colormap encodes the relative occupancy density, scaled from zero (black, unexplored regions) to unity (white/yellow, highly populated orientational states). The distinctly non-uniform spatial distribution, characterized by intense localized hotspots, geometrically confirms the clustered, meta-stable temporal dynamics driven by the non-Gaussian intermittent rotational cascade.}
	\label{fig:probability_heatmap}
\end{figure}

Figure \ref{fig:probability_heatmap} illustrates the resulting $3$D probability density heatmap $P(\mathbf{B})$ plotted over the $S^{2}$ topology. A strictly homogeneous fractional Brownian motion would progressively diffuse and isotropically cover the available geometric space. In stark contrast, the generated rotational p-model field exhibits a highly non-uniform, heavily clustered spatial distribution. The presence of sharply defined, high-density ``hotspots'' physically demonstrates that the magnetic field spends extended temporal durations trapped in localized meta-stable orientational states. These periods of quasi-stagnation are separated by the rapid, large-amplitude intermittent angular bursts analyzed in \S{} \ref{subsec:waiting_times}. Consequently, this non-ergodic macroscopic spatial clustering serves as the ultimate geometric manifestation of the underlying multifractal time-domain intermittency. Additionally, this behavior provides a clear geometric representation of anomalous transport and Lévy-flight-like dynamics in the solar wind turbulence. As extensively discussed by \citet{2004AnGeo..22.3751B}, the intermittent solar wind magnetic field can be interpreted as a competition between random-phase MHD fluctuations and coherent structures, where particles or field-line trajectories experience extended trapping inside local structures (the hotspots) punctuated by sudden, large-amplitude directional jumps.

\section{Conclusions}
\label{sec:conclusions}
This work introduced a novel, geometrically constrained stochastic framework, the $SO(3)$ rotational p-model, designed to accurately synthesize fully developed, intermittent Alfv\'enic turbulence. By rigorously mapping the singular multifractal measure of a discrete multiplicative cascade into the Lie algebra of skew-symmetric matrices $\mathfrak{so}(3)$, and subsequently integrating the phase space trajectory via a Stratonovich stochastic differential equation, the proposed methodology successfully confines the magnetic field evolution strictly onto the $S^{2}$ spherical manifold. This exact topological formulation definitively resolves the spurious variance anomalies inherent in classical Euclidean additive models, guaranteeing the absolute macroscopic conservation of the magnetic field modulus ($\delta(|\mathbf{B}|^{2})=0$) and thereby enforcing the strict kinetic incompressibility demanded by ideal Alfv\'enic dynamics.

A fundamental methodological innovation presented herein is the application of a spectral coloring operator in the frequency domain prior to the Lie-group mapping. This technique seamlessly bridges the microscopic non-Gaussian intermittency of the p-model with the macroscopic fractional scaling observed in astrophysical plasmas. The numerical experiments robustly validated this mathematical architecture: the synthesized purely directional cascade accurately reproduced the broken power-law spectral density ($f^{-1}$ to $f^{-5/3}$), the anomalous convex multifractal scaling of the structure functions, and the scale-dependent leptokurtic transition in the probability density of the projected vector increments. Furthermore, the exact geometric isolation of angular deflections allowed an unambiguous thresholding of orientational bursts, revealing a strict power-law memory in the waiting time distribution, fundamentally distinct from memoryless Poissonian processes.

When contextualized within the landscape of existing synthetic frameworks, the $SO(3)$ rotational p-model successfully overcomes the long-standing trade-offs that limited previous analytical attempts. While earlier models could only minimize compressive fluctuations at the cost of introducing unphysical local discontinuities \citep{2012PhRvL.109w1102R,2019ApJ...881L...5V,2026ApJ..1007L..18T}, or required amplification mechanisms that ultimately developed sharp structural gradients \citep{2022JPlPh..88e1703S}, this framework enforces strict mathematical incompressibility ($\delta(|\mathbf{B}|^{2})=0$) while preserving regular spatial continuity. More crucially, unlike the smooth amplification approach of \citet{2022JPlPh..88e1703S} or the localized subsets of \citet{2026ApJ..1007L..18T}, which lacked the ability to fully replicate the core statistical properties of solar wind turbulence, the methodology presented here simultaneously reproduces both the macroscopic spectral scaling (the $f^{-1}$ to $f^{-5/3}$ break) and the micro-scale multifractal intermittency. By bridging the topological rigor of the Lie algebra with the intermittent cascade of the p-model, this framework provides the first continuous, large-amplitude Alfv\'enic field that is both physically sound and statistically consistent with heliospheric observations.

As a future perspective and advanced data-driven application, the strict topological constraints of this framework can be relaxed to model phenomenological regimes exhibiting partial compressibility, such as the slow solar wind or the turbulent magnetosheath. This can be mathematically achieved by introducing a ``topological blending equation''. Let $\mathbf{B}_{rot}(t)\in SO(3)$ denote the strictly rotational, incompressible field generated by the proposed SDE, and let $\mathbf{B}_{add}(t)\in\mathbb{R}^{3}$ denote a standard Euclidean additive fluctuating field characterized by longitudinal variations. The fully observable macroscopic field $\mathbf{B}_{obs}(t)$ can be synthesized via a linear convex combination:

\begin{equation}
	\mathbf{B}_{obs}(t)=(1-C_{B})\mathbf{B}_{rot}(t)+C_{B}\mathbf{B}_{add}(t),
\end{equation}

\noindent where $C_{B}\in[0,1]$ acts as the fundamental magnetic compressibility parameter, defined as the ratio between the variance of the magnetic field magnitude and the total variance of the vector field fluctuations:

\begin{equation}
	C_{B}=\frac{E_{|\mathbf{B}|}}{E_{B_{x}}+E_{B_{y}}+E_{B_{z}}},
\end{equation}

\noindent where $E_{|\mathbf{B}|}(f)$ is the PSD of the field magnitude, and the denominator represents the trace of the spectral tensor.

When $C_{B}\to0$, the system collapses to the pure $SO(3)$ Alfv\'enic state described in this work; conversely, higher values of $C_{B}$ systematically inject controlled amounts of compressive magneto-acoustic variance. This blending paradigm provides a highly tunable, rigorous analytical toolkit for future investigations into the precise partition between directional and compressive turbulent energy across diverse heliospheric environments.

\bibliographystyle{elsarticle-harv}
\bibliography{reference_list}

\end{document}